\documentstyle[12pt,russian]{article}
\makeatletter

\newcommand{\rot}{{\rm\, rot\, }}

\renewcommand{\@cite}[2]{}
\renewcommand{\@biblabel}[1]{\hfill}

\input epsf

\begin{document}       
\large
\vspace{4cm}

PACS \quad  04.40 - b, 04.40 - Dg, 97.60 - Gb, 97.60 - Jd
$$ $$
\begin{center}
\large
{\bf General-Relativistic Curvature of Pulsar Vortex Structure.}
\end{center}

\begin{center}
\large
Shatskiy A.A.\\
Astro Space Center (ASC) Physical Lebedev Institute, Moscow, Russia.
\footnote{E-mail address for contacts: 
shatskiy@lukash.asc.rssi.ru , \,\quad
tel. in ASC: (095) 333-3366.}
\end{center}

\begin{center}
\large
Abstract --- The motion of a neutron superfluid condensate in a 
pulsar is studied. Several theorems of general-relativistic 
hydrodynamics are proved for a superfluid. The average density 
distribution of vortex lines in pulsars and their 
general-relativistic curvature are derived. 
\end{center}


Received July 3, 2000; in final form, October 6, 2000

\section{Introduction.}
\label{iiS1}

The superfluidity of matter in rotating neutron stars combines 
with the closeness of their gravitational ($r_g = 2GM/c^2$) and 
geometrical ($R$) radii, where ($G$) is the gravitational constant, 
($c$) is the speed of light, and ($M$) is the neutron-star mass 
(see, e.g., Manchester and Taylor 1980). Therefore, 
general-relativistic effects can appreciably affect the 
processes in the superfluid cores of pulsars and the mechanisms 
of glitches. In this regard, Andreev et al. (1995) discussed low 
angular velocities in general relativity: 
$\Omega < \Omega_c < \hbar /(m^*R^2)$, 
where $m^*$ is the mass of a superfluid condensate particle 
(a Cooper pair), $\Omega \sim \Omega_c$ 
(see Kirzhnits and Yudin 1995), and the number of 
vortex lines (VLs) is 0 and 1, respectively.

Here, we deal with the realistic, opposite case,
\begin{equation}
\Omega_c\ll\Omega < c/R,
\label{ii1}
\end{equation}
when there is a dense system of VLs. Recently, the interest 
in this problem has risen dramatically, which is most likely 
attributable to an increase in the accuracy of measuring glitches 
in pulsars. Consequently, it becomes possible to detect in 
principle post-Newton gravimagnetic effects in pulsars, which 
are described below. See, e.g., Prix (2000) and Langlois (2000) 
for an overview of this subject. Here, we propose a method of 
solving the problem in question that slightly differs from that 
proposed by these authors.

Bekarevich and Khalatnikov (1961) proved that rigid-body 
rotation is stable in the nonrelativistic case, and that VLs 
are rectilinear and distributed at a constant density. 
Because of general-relativistic gravielectric and gravimagnetic 
effects, which are determined, respectively, by $\nabla g_{\alpha\beta}$
 and $\rot {\bf g}$ (by the definition of Landau and Lifshitz (1988b), 
${\bf g} = \{ -g_{0\alpha}/g_{00}\}$ and $g_{ik}$ is the metric tensor), 
differential rotation becomes stable (to be more precise, 
this rotation is stable dynamically, while rigid-body 
rotation is stable kinematically as before; see below). 
In this case, VLs curve and are redistributed in space. 
These effects are described below.

\section{Analyzing changes in the system whenan ordinary fluid 
is replaced with a quantum, fermi superfluid.}

\subsection{General Relations for a Superfluid Condensate in General 
Relativity}

The wave function of a superfluid condensate is known 
(see, e.g., Bekarevich and Khalatnikov 1961) to be
$$\psi=\nu\exp(i\phi ).$$
Given the identity $u_i u^i = 1$
\footnote{The indices $i, j, k, l, m, n$ run the series 
0, 1, 2, 3; the indices $\alpha , \beta , 
\gamma $ run the series 1, 2, 3.}, 
the generalization of the nonrelativistic 
relation ${\bf v} = {\hbar\over m^*}\nabla\phi$ to general relativity is
\begin{equation}
u_i={\partial_i\phi\over k},\qquad k^2=
\partial_i\phi\partial^i\phi .
\label{ii2} \end{equation}
Indeed, let us derive an expression for a 
general-relativistic superfluid 
current $j_i = n u_i$ 
\footnote{We introduce the notation in 
which the scalar densities $\varepsilon$, $p$, $\omega = p + \varepsilon$, 
and $n$ are identified with their eigenvalues 
(i.e., with the values in a commoving frame of reference), 
respectively: the energy 
density, pressure, thermal function, and density of the 
particles; $u_i$ are the components of the 4-velocity 
vector for the matter.}
with the continuity equation $j_i^{;i} = j^i_{;i} = 0$.

Let us write the Lagrangian of a superfluid 
condensate in general relativity by using the 
Madelung hydrodynamic representation (see Grib 
et al. 1980; Bogolyubov and Shirkov 1993):
\begin{equation}
L={1\over 2}\psi^*_{;i}\psi^{;i}+|\psi |^2 \tilde F(|\psi |^2)=
{1\over 2}\nu^2\partial_i\phi\partial^i\phi + \nu^2 F(\nu^2).
\label{ii19} \end{equation}
The principle of least action relative to a change in $\phi$ yields
\begin{equation}
(\nu^2\partial^i\phi )_{;i}=0.
\label{ii2a} \end{equation}
Consequently, the current vector is 
$j^i = \nu^i = const\cdot\nu^2\partial^i\phi$. 
This is seen from a comparison of $j^i$ and 
$j_{sf}^i$ in the nonrelativistic case: $u^i = v^i /c$ 
and $j^i = nv^i /c$. For a superfluid, we write in 
this case: $n = \nu^2$, $v^i = {\hbar\over m^*}\partial^i\phi$, 
and $j_{sf}^i ={\hbar\over m^* c}\nu^2\partial^i\phi$. 
Since the continuity equation $j^i_{;i} = 0$ must hold 
in any case, we derive Eq.~(4) for a superfluid. 
Hence, we obtain for a superfluid current in general relativity
\begin{equation}
j^i={\hbar\over m^* c}\nu^2\partial^i\phi = nu^i.
\label{ii3b} \end{equation}
or, for covariant quantities,
\begin{equation}
j_i={\hbar\over m^* c}\nu^2\partial_i\phi = nu_i.
\label{ii3a} \end{equation}
A scalar multiplication of the latter 
expression by Eq.~(5) yields
$$\left( {\hbar\over m^* c}\right)^2\nu^4\partial^i\phi
\partial_i\phi =n^2 ,$$
Expression (2) follows from this.

In the Newton approximation, $k^2\approx (m^* c/\hbar )^2$. 
The scalar $k$ is identified with $w/(nc\hbar )$ (see below). 
Hence, since $\nu^2 = n$ in the Newton approximation, 
Eq.~(6) or (2) yields
\begin{equation}
u_i \approx {\hbar\over m^*c}\partial_i\phi .
\label{ii4} \end{equation}
In seeking to approach an optimum regime, a 
superfluid current undergoes a well-known 
rearrangement (see Bekarevich and Khalatnikov 1961). 
As a result, while the current remains potential 
"almost everywhere", vorticity arises, 
which generally coincides with the vorticity for an 
ordinary (nonsuperfluid) fluid at the same point. 
This occurs, because a system of VLs is formed, 
with the phase of the wave function $\phi = \varphi +f(t)$ 
corresponding to each of them 
\footnote{In what follows, we use a cylindrical coordinate 
system with the $z$ axis directed along the spin axis: 
$x^{0,1,2,3} = ct,\rho ,\phi ,$ and $z$.}. According to (7), 
we then have for a single vortex
\begin{equation}
u_{\varphi} \approx {\hbar\over m^*c}.
\label{ii5} \end{equation}
On the vortex axis itself, where $\phi$ is uncertain, 
the wave function loses its meaning, and the superfluidity 
vanishes. This corresponds to a physically nonsuperfluid VL 
"core" of a macroscopically small radius, which is 
responsible for the nonzero curl of velocity (see Bekarevich 
and Khalatnikov 1961).

The above reasoning is universal and equally applies to the 
nonrelativistic and general-relativistic cases.

\subsection{The General-Relativistic Bernoulli Theorem}

For the subsequent analysis of a superfluid in general 
relativity, we need the general-relativistic Bernoulli theorem.

The classical Bernoulli theorem is derived for a steady, 
isentropic fluid flow from the Euler hydrodynamic equations 
(see Landau and Lifshitz 1988a):
$${\partial {\bf V}\over\partial t}+
({\bf V}\nabla ){\bf V}=-\nabla {w\over mn}.$$
In the relativistic case, these equations can be written as
\begin{equation}
u^i\partial_i(k^*u_j) =\partial_j k^*,
\label{ii13} \end{equation}
where $k^* = \omega /n$. Note that these equations have a particular 
solution of the form $k^* u_j = const\cdot\partial_j\phi$, 
where $\phi$ is a scalar 
function of the coordinates. Therefore, in contrast to the 
nonrelativistic case, the quantity $k^* u_j$ rather than the 
velocity has the potential. Hence, according to (2), we 
immediately identify $k^*$ for a superfluid with $c\hbar k$:
\begin{equation}
k^*={w\over n}=\hbar ck
\label{ii13.2} \end{equation}
In general relativity, $\partial_j$ is replaced with the covariant 
derivative $\nabla_j$:
\begin{equation}
u^i\nabla_i(k^*u_j) =\nabla_j k^*,
\label{ii14} \end{equation}
or, expanding the covariant derivative,
$$u^i\partial_i(k^*u_j)-u^iƒ_{ij}^mk^*u_m=\partial_j k^*.$$
Substituting the expressions for the Christoffel symbols 
in this relation,
\begin{equation}
ƒ_{ij}^m={1\over 2}g^{mn}(\partial_i g_{jn}+\partial_j g_{in}-
\partial_n g_{ij}),
\label{ii15} \end{equation}
yields
\begin{equation}
u^i\partial_i(k^*u_j)-{1\over 2}k^*u^iu^n
\partial_j g_{in}=\partial_j k^*.
\label{ii16-1} \end{equation}
Denoting
\begin{equation}
\phi_j=\partial_j\phi =ku_j ;\quad\phi^j=\partial^j\phi =ku^j,
\label{ii21} \end{equation}
we obtain for a superfluid
\begin{equation}
\phi^i\partial_i\phi_j -{1\over 2}\phi^i\phi^n
\partial_j g_{in}={1\over 2}\partial_j {k}^2.
\label{ii16-2} \end{equation}
In that case, since $k^2 = g_{in}\phi^i\phi^n$, we derive two equations,
\begin{equation}
\phi^\alpha[\partial_\alpha\phi_0-\partial_0\phi_\alpha]=0 .
\label{ii16-4} \end{equation}
and
\begin{equation}
\phi^\alpha[\partial_\alpha\phi_\gamma-\partial_\gamma\phi_\alpha]+
\phi^0[\partial_0\phi_\gamma-\partial_\gamma\phi_0]=0 .
\label{ii16-5} \end{equation}
The former follows from the latter. Denoting
\begin{equation}
\sigma_{\alpha\gamma}={1\over 2\pi}[\partial_\alpha\phi_\gamma
-\partial_\gamma\phi_\alpha ],
\label{ii16-6} \end{equation}
we obtain for $\sigma_{\alpha\gamma}$
\footnote{Since the condensate 
ceases to exist on vortex lines, the concept of phase loses 
its meaning; thus, the theorem on the curl of a gradient 
ceases to be valid on vortex cores, and the 4-gradient of 
phase $\partial_i \phi$ is replaced with a 4-vector $\phi_i$.}:
\begin{equation}
\sigma_{\alpha\gamma}=
{\phi^0\over 2\pi\phi^\beta\phi_\beta}
[\phi_\alpha\partial_\gamma\phi_0-\phi_\gamma\partial_\alpha\phi_0
+\phi_\gamma\partial_0\phi_\alpha-\phi_\alpha\partial_0\phi_\gamma].
\label{ii16-8} \end{equation}
We emphasize that our system is not axially symmetric 
because it contains vortices, but this symmetry is restored 
by averaging over the VL distribution. As a result, there is 
no steady state in this case. Clearly, an infinite number of 
stationary frames of references exist for an ordinary fluid 
uniformly rotating in a vessel. Below, we show that there 
is a (unique) frame of reference for a rotating superfluid comoving 
with VL cores in which all quantities are stationary. For 
this frame of reference, we derive from (17) $\partial_\gamma \phi'_0 = 0$ 
\footnote{The prime means that the quantities under consideration 
refer to a stationary frame of reference outside VL cores.}
on all current lines where $\sigma_{\alpha\gamma} = 0$
($\partial_\gamma$ in any direction). Therefore,
\begin{center}
\bigskip
\fbox{%
\parbox{12cm}{%
\begin{equation}
ku'_0=\phi'_0=const.
\label{ii16-9} \end{equation}
}%
}
\end{center}
This is the general-relativistic analog of the 
Bernoulli theorem. In the nonrelativistic limit 
(see Landau and Lifshitz 1988a):
$$k ={1\over\hbar c}{w\over n}\to
{m^*c\over\hbar}(1+{p\over\tilde\rho c^2}),\quad
u_0=\sqrt{{g_{00}\over 1-v^2/c^2}}\to 
(1+{\tilde\varphi\over c^2})(1+{v^2\over 2c^2})\to $$
$$\to 1+{\tilde\varphi\over c^2}+{v^2\over 2c^2},
\Longrightarrow ku_0\to {m^*c\over\hbar}\,
(1 + {\tilde\varphi + v^2/2 + p/\tilde\rho\over c^2}) = const,$$
where $\tilde\phi$ and $\tilde\rho$ are the Newton gravitational 
potential and the matter density, respectively. 
This is the ordinary Bernoulli theorem.

For a rotating superfluid with vortices, the laboratory 
frame of reference ceases to be stationary, but we can 
choose a frame of reference for a rotating superfluid that 
comoves with VL cores; in this case, the frame of reference 
is stationary and does not comove with the superfluid.

\section{Determining the relationship between dynamical 
superfluid parameters}
\subsection{The Principle of Least Action for a 
Rotating Superfluid in General Relativity}

Hartl and Sharp (1967) proved that rigid-body rotation 
with an angular velocity equal to the shell angular 
velocity is most favorable for an ideal, ordinary fluid 
in general relativity.

Let us prove that this assertion remains also 
valid for a superfluid
\footnote{By the angular 
velocity of a superfluid, we imply the mean value 
of $u^\varphi /(cu^0)$ averaged over the surface orthogonal to the vortex 
direction near the point under consideration whose area is 
much smaller than the system's cross-sectional area, but, 
at the same time, is still much larger than the square 
of the mean separation between vortices. As we show 
below, this condition can definitely be satisfied for pulsars.}.

In view of the general-relativistic Bernoulli theorem and 
by analogy with the study of Hartl and Sharp (1967) for 
an ordinary fluid, we assume the spatial components $\partial_\gamma\phi$ 
and the probability density $\nu^2$ for a condensate particle 
to be detected at a given point to be independent variables 
for superfluid dynamics. In addition, we will remember 
that $x_k$ are the variables that determine the vortex shape 
and coordinates, while the metric components $g_{ik}$ are the 
variables that determine the system's gravitational field.

Given the conservation of momentum and the total number 
of particles, the expression for the total energy of a 
superfluid can be written as (Hartl and Sharp 1967)
\begin{equation}
\tilde L=\nu^2\phi_0\phi^0 -\nu^2\, [F+k^2/2]
-\Omega (-{1\over c}\nu^2\phi^0\phi_\varphi )-\mu\nu^2\phi^0 ,
\label{ii19-3} \end{equation}
where $\Omega$ and $\mu$ are the corresponding Lagrange factors.

In addition, we must take into account the Lagrangian 
$L_0$ of the shell. Given momentum conservation, the 
analog of (21) for the shell can be written as
$$\tilde L_o=J_o\Omega_o^2/2 - \Omega J_o\Omega_o,$$
where $J_0$ and $\Omega_0$ are the moment of inertia and angular velocity 
of the shell in the frame of reference under consideration, 
respectively. Varying over $\Omega_0$ leads to the equality
$$\Omega =\Omega_o.$$
If, however, the satisfaction of the general-relativistic 
potentiality condition for a superfluid flow must also be 
taken into account, then, according to the general rules 
for imposing additional conditions, we must add the 
term $\Delta\tilde L$ to $\tilde L$ to express the fact that the circulation 
of the phase gradient over a closed contour is proportional 
to the number of vortices crossing this contour. This 
is the general-relativistic generalization of the potentiality 
condition for a superfluid flow:
\begin{equation}
\oint\limits_ƒ \partial_\gamma\phi\, dx^\gamma =\int\limits_ƒ
(\partial_\beta\partial_\gamma\phi
-\partial_\gamma\partial_\beta\phi )\, dx^\beta\land dx^\gamma =2\pi K,
\label{ii20-6} \end{equation}
where $dx^\beta\wedge dx^\gamma$ is the directed surface element pulled over 
contour $\Gamma$, and $K$ is the number of vortices crossing 
the contour. Thus, the general-relativistic potentiality 
condition, according to (18), can be written as
\begin{equation}
2\pi\sigma_{\beta\gamma}=\sum_{k=1}^N
2\pi n^k_{\beta\gamma}\delta^2(x-x_k),
\label{ii20-11} \end{equation}
while the addition to $\tilde L$ corresponding to this condition can 
be written as
\begin{equation}
\Delta\tilde L=\xi^{\beta\gamma}(x)
[2\pi\sigma_{\beta\gamma}-\sum_{k=1}^N
2\pi n^k_{\beta\gamma}\delta^2(x-x_k)].
\label{ii20-7} \end{equation}
Here, according to (18), $2\pi\sigma_{\beta\gamma}$ 
is the curl of $\phi_\gamma$; $n^k_{\beta\gamma}$ 
is a unit antisymmetric tensor dual to one of the surfaces, 
which are orthogonal to the vectors of the vortex direction 
at each point, $n^k_{\beta\gamma}$ is defined at the point of intersection 
of this surface with vortex $k$, and the summation over $k$ is 
performed over all vortices in the system
\footnote{In general, addition (24) to $\tilde L$ 
should be written for each such 
surface, but this is implied by default.};
$\delta^2 (x-x_k)$ is 
the bivariate delta function, and $x$ is the radius vector 
defined on this surface; and the antisymmetric tensor $\xi^{\beta\gamma}(x)$ 
is the Lagrange functional factor.

\section{Determining the Relationship between the Components of 
the 4-Gradient of Phase for a Superfluid Condensate}

Action $A$ in general relativity is known to be related to 
Lagrangian $L$ by
$$A=\int L\, dX,\quad dX=\sqrt{-g}\, d^4x,$$
where the integration is performed over 4-space; $g$ is the 
determinant of the metric tensor $g_{ik}$. We denote the integral 
of $\tilde L$ over space by
\begin{equation}
\tilde E=\int\sqrt{-g}\tilde L\, d^3x,\qquad
(\, \Delta\tilde E=\int\sqrt{-g}\Delta\tilde L\, d^3x\, ).
\label{ii20-9} \end{equation}
By its physical meaning, $\tilde E$ is the system's energy for 
imposed additional conditions, such as allowance for the 
conservation of angular momentum, the total number of 
particles in the system, etc.

We break up the integral over space into an integral over the 
surface dual to the tensor $n^k_{\beta\gamma}$ and integrals over the lengths 
of the vortices orthogonally crossing this surface. Because 
of the presence of delta functions, the following sum remains 
from $\Delta\tilde E$:
$$\Delta\tilde E=\int\sqrt{-g}
2\xi^{\beta\gamma}(x)(\partial_\beta\phi_\gamma )d^3x +
\sum_{k=1}^N 2\pi l_k\sqrt{-g}
\xi^{\beta\gamma}(x_k)n^k_{\beta\gamma},$$
where $l_k$ is the length of vortex line $k$.

After varying $\Delta\tilde E$ over $\phi_\gamma$, the following term remains:
\begin{equation}
-2\partial_\beta(\sqrt{-g}\xi^{\beta\gamma})
\label{ii20-5} \end{equation}
If we discard addition (26), then the subsequent analysis 
will be valid only for an averaged description of the 
superfluid motion. Let us prove that term (26) vanishes on 
vortex cores. Since the coordinates $x_k$ must correspond to 
equilibrium vortex positions in the system, the system must 
be stable against core displacements orthogonal to the 
direction of the vortices themselves:
\begin{equation}
2\pi l_kn^k_{\beta\gamma}
[\partial_\beta(\sqrt{-g}\xi^{\beta\gamma})]|_{x=x_k}=0.
\label{ii20-8} \end{equation}
This proves the above assertion. The corrections related 
to the term $\Delta\tilde E$ in action will not be considered 
everywhere, because we are interested in the system's 
dynamics only when expression (27) holds, i.e., on vortex 
cores, or, equivalently, only an averaged description of all 
quantities for the system. Below, we thus denote average 
quantities by a hat above the symbol (e.g., $\hat A$). According 
to Eq.~(3), the principle of least action relative to a 
change in $\nu^2$ leads to the equation
\begin{equation}
{\partial L\over\partial \nu^2}={\partial\over
\partial\nu^2}\{\nu^2\, [k^2/2 + F(\nu^2)]\}=0.
\label{ii19-5} \end{equation}
Given that $k^2 =  g^{\alpha\beta}\phi_\alpha \phi_\beta +2g^{0\alpha}
\phi_0 \phi_\alpha +g^{00}(\phi_0 )^2$
and denoting
\begin{equation}
A^\alpha ={\partial\phi_0\over\partial\phi_\alpha},\qquad
B^\alpha ={\partial\phi^0\over\partial\phi_\alpha},
\label{ii2-2} \end{equation}
we obtain
\begin{equation}
{\partial k^2\over\partial\phi_\alpha}=2g^{\alpha\beta}\phi_\beta
+2g^{0\alpha}\phi_0 + 2g^{0\alpha}\phi_\alpha A^\alpha +2g^{00}\phi_0
A^\alpha = 2[\phi^\alpha + \phi^0 A^\alpha ].
\label{ii19-7} \end{equation}
Taking a variational derivative of Eq.~(21) with respect 
to $\phi_\gamma$ and $\nu^2$,
using Eqs.~(28), (29), (30), we derive for 
the average quantities
\begin{equation}
\hat B^\gamma
(\hat\phi_0 +{\Omega\over c}\hat\phi_\varphi -\mu )=
\hat\phi^\gamma - {\Omega\over c}\hat\phi^0\delta^\gamma_\varphi .
\label{ii24} \end{equation}
\begin{equation}
\hat\phi_0 +{\Omega\over c}\hat\phi_\varphi -\mu =0.
\label{ii25} \end{equation}
Comparing Eqs.~(31) and (32), we obtain the analog of 
rigid-body rotation for a superfluid:\\
\begin{center}
\bigskip
\fbox{%
\parbox{12cm}{%
\begin{equation}
{\phi^\gamma\over\phi^0}|_{x=x_k} =
{\hat\phi^\gamma\over\hat\phi^0}={\hat{d x^\gamma\over d x^0}}
={\Omega\over c}\delta^\gamma_\varphi .
\label{ii26} \end{equation}
}%
}
\end{center}
\bigskip

\subsection{Magnus Force and the General-Relativistic Theorem on 
the Conservation of Circulation}

The Magnus force acts on a rotating body in an incoming flow 
and is attributable to a nonzero pressure difference for the 
opposite sides of the flow around the body. In tern, the 
pressure along the body boundary changes because of the Bernoulli 
theorem: the velocity of the medium that flows around a 
rotating body changes when going around the body axis.

Let us write the Bernoulli equation for the 
nonrelativistic case:
\begin{equation}
p+mn {\bf V}^2/2 =const
\label{iiP-1} \end{equation}
It would be natural to consider a portion of the VL core as 
the body on which the Magnus force acts. As we 
show below, the Magnus force does not depend on the 
radius $a$ of this core. Choose a cylindrical coordinate 
system whose $z$ axis coincides with the rotation axis of this core. 
Since the core radius $a$ is much smaller than any scales 
on which the incoming flow 
produced by the remaining vortices changes appreciably, 
this flow may be considered constant for 
the flow around the core. According to (33), the physical 
velocity of this flow at the core point is
$${\bf V}_0 = [{\bf \Omega\times r}]$$
At the same time, the velocity produced by the core itself 
on its boundary, according to (8) is
$${\bf V}_k = {\hbar\over m^*a^2}[{\bf e}_z\times {\bf a}],$$
where ${\bf a}$ and ${\bf e_z}$ are the vector in the direction of the core 
radius (equal to $a$ in magnitude) and a unit vector along 
the core rotation axis, respectively. The total velocity 
on the core boundary is
$${\bf V} = {\bf V}_0 + {\hbar\over m^*a^2}[{\bf e}_z\times {\bf a}].$$
The force per unit core area is equal to the pressure 
multiplied by $a$ unit vector: $-{\bf e_a} = -{\bf a}/a$. Accordingly, 
the force per unit core length is
\begin{equation}
{\bf F} = -\oint{\bf e}_a p(\varphi )\, dl=
-\oint\limits_0^{2\pi}{\bf e}_a p(\varphi )a\, d\varphi .
\label{iiP-2} \end{equation}
Expressing $p$ from Eq.~(34) and substituting it in Eq.~(35) yield
\begin{equation}
{\bf F} = -\oint\limits_0^{2\pi}{\bf a}\{ const - 
mn({\bf V}_0^2+{\bf V}_k^2 + 2{\bf V}_0 {\bf V}_k)/2\}d\varphi .
\label{iiP-3} \end{equation}
Since the $\varphi$ --- independent terms vanish and since $n\approx\nu^2$ 
and $m\approx m^*$ for a superfluid in the nonrelativistic case, 
we derive for the Magnus force per unit core length
\begin{equation}
{\bf F} = \oint\limits_0^{2\pi}{\bf a}\{mn{\bf V}_0
{\hbar\over m^*a^2}[{\bf e}_z\times {\bf a}]\}d\varphi =
\pi\hbar\nu^2 [{\bf V}_0\times {\bf e}_z] .
\label{iiP-3} \end{equation}
It follows from Eq.~(37) that the Magnus force acting on the 
VL core that is at rest relative to a remote observer causes 
it to move toward the vessel wall. As it accelerates toward 
the wall, an incoming flow emerges (to be more precise, the 
core itself runs on the superfluid); as a result, the Magnus 
force changes its direction, causing the VL core to precess 
around some point of the superfluid flow. This precession is 
rapidly damped, and the core starts moving in such a way that 
the Magnus force does not act on it, i.e., that the system's 
energy becomes minimal. This requires that the core be at rest 
with respect to the superfluid flow at its location. The above 
reasoning proves that the VL system is frozen in, i.e., that 
there is no slip in the nonrelativistic case.   

In general relativity, expression (34) for the Bernoulli theorem 
is replaced with expression (20).

For the nonrelativistic case, Hess (1967) showed that the VLs 
are, as it were, frozen in a superfluid --- move at angular 
velocity $\Omega$, the shell rotation velocity. Thus, there 
is no slip relative to this angular velocity.

To generalize the slip theorem to general relativity, we do not 
need to repeat similar calculations in order to determine the 
Magnus force, which, incidentally, are very complex. It will 
suffice to note that $\phi_0 = ku_0$, $k = k(|{\bf V}|,p,n,g_{ik})$, 
and $u_0 = u_0(|{\bf V}|,g_{ik})$ and that 
when going around the core of a vortex line, the changes in 
metric $g_{ik}$ and density $n$, if any, are so negligibly small 
\footnote{The core radius is known to have sizes of the order 
of $1\div 10$ interparticle separations.} 
that they may be disregarded. 

Therefore, we can write for the superfluid portion adjacent to 
the core
\begin{equation}
p=p(|{\bf V}|)
\label{iiP-4} \end{equation}
We thus see that there is no general-relativistic Magnus 
force for a VL motion with $|{\bf V}| = const$ around the core. 
This is possible only when the VL core accompanies the 
superfluid flow, i.e., when ${\bf V}$ is produced by the core itself.

Consequently, given Eq.~(33), we see that there is no slip 
in general relativity either.

In conclusion, note that the absence of slip also follows 
from another important theorem of hydrodynamics, the theorem 
on the conservation of circulation (see, e.g., Landau and 
Lifshitz 1988a). 

\section{Calculating the mean density and curvature of vortex 
lines in a pulsar with general-relativistic corrections}

\subsection{Passing to a Rotating Frame of Reference}

As will be evident below, it is more convenient to perform 
an analysis in a comoving (with vortex cores), i.e., rotating 
frame of reference.

According to Landau and Lifshitz (1988b), the following 
formulas can be derived that relate the tensor components 
in various frames of reference:
\begin{equation}
\left\{
\begin{array}{rcl}
g'_{\rho\rho}=g_{\rho\rho};\quad g'_{\varphi\varphi}=g_{\varphi\varphi};
\quad g'_{zz}=g_{zz}\\
g'_{0\varphi}=g_{0\varphi}+{\Omega\over c}g_{\varphi\varphi};\quad
g'_{00}=g_{00}+({\Omega\over c})^2g_{\varphi\varphi}+
2{\Omega\over c}g_{0\varphi}\\
u'_\rho =u_\rho ;\quad u'_\varphi =u_\varphi ;\quad u'_z=u_z;
\quad u'_0=u_0+{\Omega\over c}u_\varphi .\\
\end{array}\right.
\label{ii32} \end{equation}
Here, the components in a frame of references rotating with 
angular velocity $\Omega$ are marked by primes.

\subsection{Calculating the Covariant Curl of Superfluid Velocity 
and the Vortex Density in the System}

By definition, the mean density of vortices is the number of 
vortices crossing the orthogonal surface divided by its area. 
Therefore, using Eq.~(22) to derive the vortex density 
$\sigma_{\beta\gamma}$, we obtain
\begin{equation}
\sigma_{\beta\gamma}={K\over dx^\beta\land dx^\gamma}=
{1\over 2\pi}(\partial_\beta\phi_\gamma
-\partial_\gamma\phi_\beta ).
\label{ii33} \end{equation}
It is convenient to express the quantities $\phi_\gamma$ in terms 
of $\phi_0$, the metric, and the shell angular velocity in 
the frame of reference under consideration:
$$\left\{
\begin{array}{rcl}
\phi_\alpha =\phi^0 g_{0\alpha}+\phi^\gamma g_{\alpha\gamma}=
\phi^0(g_{0\alpha}+g_{\alpha\gamma}{\phi^\gamma\over\phi^0}),\\
\phi_0=\phi^0 g_{00}+\phi^\gamma g_{0\gamma} =
\phi^0(g_{00}+g_{0\gamma}{\phi^\gamma\over\phi^0}).\\
\end{array}\right. $$
Hence, according to~(33), we obtain for the average quantities
\begin{equation}
\hat\phi_\alpha =\hat\phi_0{g_{0\alpha}+{\Omega\over c}
g_{\varphi\alpha}\over g_{00}+{\Omega\over c}g_{\varphi 0}}.
\label{ii34} \end{equation}
Since the average quantities do not depend on time and 
angle $\varphi$, we introduce the notation
\begin{equation}
X_\gamma =2\pi\hat\sigma_{\varphi\gamma}=-\partial_\gamma\hat\phi_\varphi ;
\quad Y=\hat\phi_0 ;\quad Z=
-{g_{0\varphi}+{\Omega\over c}g_{\varphi\varphi}\over
g_{00}+{\Omega\over c}g_{\varphi 0}},
\label{ii35} \end{equation}
and derive from~(41)
\begin{equation}
X_\gamma =\partial_\gamma (YZ).
\label{ii36} \end{equation}
\nopagebreak[4]
On the other hand, we obtain from Eqs.~(19) and (33)
\nopagebreak[4]
\begin{equation}
X_\gamma =(c/\Omega)\partial_\gamma Y ,
\label{ii37} \end{equation}
Solving the last two equations for $X_\gamma$ and $Y$ yields
\begin{equation}
X_\gamma =const\, {\partial_\gamma Z\over (1-{\Omega\over c}Z)^2},
\label{ii38} \end{equation}
\begin{equation}
Y ={const\over 1-{\Omega\over c}Z}.
\label{ii39} \end{equation}
Since $Y = \hat\phi_0 \to k \to m^*c/\hbar$ 
in the nonrelativistic limit, we see that $const = m^*c/\hbar$.
Hence, we have
\begin{equation}
\hat\sigma_{\varphi\gamma} ={m^*c\over 2\pi\hbar }
{\partial_\gamma Z\over (1-{\Omega\over c}Z)^2}.
\label{ii40} \end{equation}
As was shown above, $\Omega' = 0$ in the comoving (with cores) 
frame of reference; therefore, Eq.~(47) in this frame of 
references is especially simple:
\begin{center}
\bigskip
\fbox{%
\parbox{12cm}{%
\begin{equation}
{\hat{\sigma^\alpha}}' = {m^*c\over 2\pi\hbar\sqrt{\tilde\gamma'}}
e^{\alpha\gamma\varphi}\partial_\gamma g'_{\varphi},
\label{ii41} \end{equation}
}%
}
\end{center}
where, according to Landau and Lifshitz (1988b), 
$g_\gamma = -g_{0\gamma}/g_{00}$, $\tilde\gamma = -g/g_{00}$ 
is the determinant 
of the spatial metric tensor, and the three-dimensional 
vector $\sigma^alpha$ dual to the tensor $\sigma_{\gamma\beta}$
was defined as $\sigma^\alpha = 
(2\sqrt{\tilde\gamma})^{-1}\cdot e^{\alpha\gamma\beta}\sigma_{\gamma\beta}$,
$e^{\alpha\gamma\beta}$ is a unit antisymmetric tensor.
The vector $\hat\sigma^\alpha$ coincides in direction with the vortex 
direction in the system and is equal in magnitude to the 
mean vortex density at a given point.

For the invariant mean vortex density, we can write
\begin{equation}
\hat\sigma =\sqrt{\hat\sigma_{ij}\hat\sigma^{ij}}=
\sqrt{2\hat\sigma_{0\alpha}\hat\sigma^{0\alpha}+\hat\sigma_{\alpha\beta}
\hat\sigma^{\alpha\beta}}
\label{iidopol1} \end{equation}
As a result, we have for the invariant density in the first 
post-Newton approximation
\begin{equation}
\hat\sigma \approx |\hat\sigma_{\varphi\alpha}|
\sqrt{g^{\alpha\alpha}g^{\varphi\varphi}},
\label{iidopol2} \end{equation}
which matches the nonrelativistic limit for the VL density.

According to Eqs.~(39), we have for $g'_\alpha$
\begin{equation}
g'_\alpha =\delta^\varphi_\alpha{g_\varphi -{\Omega\over c}
{g_{\varphi\varphi}\over g_{00}}\over 1+({\Omega\over c})^2
{g_{\varphi\varphi}\over g_{00}}-2g_\varphi {\Omega\over c}}.
\label{ii36} \end{equation}
Given that $g' = g$, we obtain for $\tilde\gamma'$:
$$\tilde\gamma' =-{g'\over g'_{00}}={\tilde\gamma\over 1+({\Omega\over c})^2
{g_{\varphi\varphi}\over g_{00}}-2{\Omega\over c}g_{0\varphi}}.$$
In the nonrelativistic case, $\tilde\gamma \to\rho^2$, 
$g_{00}\to 1$, $g_{\varphi\varphi}\to -\rho^2$, 
$g_{zz}\to -1$, $g_{0\gamma}\to 0$, and $\phi_0 \to k\to m^*c/\hbar$; 
therefore, we derive the already 
known expression in this limit
$$\hat\sigma^\alpha\to\hat\sigma^\alpha_0=
\delta^\alpha_z{\Omega m^*\over\pi\hbar}=const.$$
For a millisecond pulsar, $\hat\sigma_0^\alpha \sim 10^{-6}\, cm^{-2}$, 
with a separation between vortices $d\sim 10^{-3}\, cm$ corresponding 
to this value. As for the radius of the core itself, its 
order-of-magnitude value is $10^{-11}\, cm$.

As we see from this section, the general-relativistic 
corrections that determine the curvature of VLs and the 
change in their mean density are small for real pulsars, 
and, therefore, the relative curvature does not exceed 
a few percent. 

\section{Calculating corrections for a homogeneous model.}

Let us calculate the invariant density of vortex lines 
in a pulsar in the first post-Newton approximation. The 
model is based on the assumption that the pulsar interiors 
rotate at angular velocity $\Omega$ and, because the 
compressibility of a neutron condensate is low, its density 
is assumed to be $\tilde\rho$ in the entire volume
\footnote{For convenience of calculations, we take $G = 1$ and $c = 1$
in this section.)}.

In the first post-Newton approximation, the metric in a 
conformally Euclidean coordinate system
\footnote{It is easy to see that the result does not depend on the 
choice of a coordinate system in this approximation.)}
can be written as
\begin{equation}
ds^2=(1+2\tilde\varphi)dt^2 -(1-2\tilde\varphi)
[d\rho^2+dz^2+\rho^2 d\varphi^2]-2g_\varphi dt\, d\varphi .
\label{iip1}\end{equation}
According to Eq.~(50), we write
\begin{equation}
\hat\sigma =\sqrt{g^{\gamma\gamma}g^{\varphi\varphi}}
\hat {\sigma_0\over 2\Omega}\partial_\gamma [g_\varphi -\Omega
g_{\varphi\varphi}/g_{00}],\qquad (\hat\sigma_0={\Omega m^*\over \pi\hbar}).
\label{iip2} \end{equation}
Since $g^{\alpha\alpha} = 1/g_{\alpha\alpha}$, 
the corrections to $\hat\sigma_0$ can be arbitrarily 
divided up into two groups:

(1) Gravimagnetic corrections: 
$$\hat\sigma_1=
({m^*\over 2\pi\hbar}){(\partial_\rho +\partial_z)\, g_\varphi\over\rho}.$$

(2) Gravielectric corrections: 
$$\hat\sigma_2={\hat\sigma_0\over 2}{1+2\tilde\varphi\over\rho}
(\partial_\rho +\partial_z )
[\rho^2{1-2\tilde\varphi\over 1+2\tilde\varphi }]-\hat\sigma_0 .$$
We thus see that the VL curvature and redistribution in a 
pulsar result from the gravimagnetic interaction of VLs with 
the Lense-Tirring field of the system and from the gravielectric 
deformation of the system's Euclidean geometry.

The Newton gravitational potential $\tilde\varphi$ of the model can be 
easily calculated:
\begin{equation}
\tilde\varphi =-2\pi\tilde\rho R^2(1-x^2/3-y^2/3)\qquad (x=\rho /R,\,\,
y=z/R).
\label{iip3}\end{equation}
Hence, it is easy to derive an expression for the gravielectric 
corrections:
$$
\hat\sigma_2=4\pi\tilde\rho R^2\hat\sigma_0(1-x^2-y^2/3-2xy/3).
$$
Given that $\tilde\varphi_{_R} = -4\pi\tilde\rho R^2/3$ 
on the stellar surface, we obtain for the 
gravielectric corrections
\begin{equation}
\hat\sigma_2(x,y)=3|\tilde\varphi_{_R}|\hat\sigma_0(1-x^2-y^2/3-2xy/3).
\label{iip4}\end{equation}
To determine the gravimagnetic corrections, we use Eq.~(106.15) 
from Landau and Lifshitz (1988b) to derive the metric 
components $g_\varphi$. In Cartesian coordinates,
\begin{equation}
g_{0\alpha}({\bf r})={1\over 2}\int\limits_V \tilde\rho\, dr'^3\{  
{7[{\bf\Omega\times r'}]_\alpha +
([{\bf\Omega\times r'}]_\beta\, , n_\beta )\, n_\alpha\over |{\bf r-r'}|}\} ,
\label{iip5}\end{equation}
where $n_\alpha = (r_\alpha -r'_\alpha )/|{\bf r-r'}|$.

Hence, it is easy to calculate the angular component of 
the gravimagnetic field in cylindrical coordinates:
\begin{equation}
g_{\varphi}({\bf r})=-{\rho\over 2}\int\limits_V \tilde\rho\, 
d\varphi |_{-\pi}^{+ \pi} dz'\rho' d\rho'\{  
{7\Omega\rho'\cos\varphi\over |{\bf r-r'}|} +
{\Omega\rho' p^2\cos\alpha\cos\gamma\over |{\bf r-r'}|^3}\} ,
\label{iip6}\end{equation} 
Here, 
$({\bf r-r'})^2 =(z-z')^2+p^2,\quad p^2=\rho^2+{\rho'}^2-
2\rho\rho'\cos\varphi $,\\
$\cos\alpha =\cos\varphi\sin\theta + \sin\varphi\cos\theta ,\quad$
$\cos\gamma =-\sin\theta ,\quad\sin\theta ={\rho'\over p}\sin\varphi ,\quad$
and $\cos\theta ={p^2+\rho^2 -{\rho'}^2\over 2p\rho }$. \\
As a result, expression~(57) reduces to
\begin{equation}
g_{\varphi}({\bf r})=-{\Omega\rho\over 2}\int\limits_V \tilde\rho\, 
d\varphi |_{-\pi}^{+ \pi} dz'\rho' d\rho'\{  
{7\rho'\cos\varphi\over |{\bf r-r'}|} +
{\rho'\sin^2\varphi (\rho'^2+p\rho -p\rho'\cos\varphi )\over |{\bf r-r'}|^3}\}.
\label{iip7}\end{equation} 
Since the integrand is even in variable $\varphi$, by changing 
variables: $x = \rho /R$, $x' = \rho' /R$, $y=z/R$, 
$y'(y) =(z'-z)/R$, and 
$p'(x,\varphi ) = p/R$, we can write the expression for the 
gravimagnetic corrections as
\begin{equation}
\hat\sigma_1(x,y)=\hat\sigma_0 |\tilde\varphi_{_R}|
\int\limits_{0}^{\pi}\, d\varphi
\int\limits_0^1  dx'
f(x,x',y,\varphi )
\label{iip8}\end{equation} 
where
\begin{equation}
\begin{array}{lll}
f(x,x',y,\varphi )=-{3x'^2\over 8\pi x}(\partial_x +\partial_y )
\int\limits_{y'_1}^{y'_2}\, dy'\left\{
{7x\cos\varphi\over [y'^2+p'^2]^{1/2} }+
{xx'\sin^2\varphi (x'^2+p'x -p'x'\cos\varphi )\over 
[y'^2+p'^2]^{3/2}} \right\} ,\\
\, \\
y'_1=-\sqrt{1-x'^2}-y,\quad y'_2=+\sqrt{1-x'^2}-y.\\
\end{array}
\label{iip9}\end{equation}
Hence, integrating and then differentiating yields
$$ 
\begin{array}{lll}
A=7x\cos\varphi ,\\
B=xx'\sin^2\varphi (x'^2+p'x-p'x'\cos\varphi ) ,\\
C=\sqrt{y'^2+p'^2}\\
A_x=7\cos\varphi ,\\
B_x=x'\sin^2\varphi (x'^2+p'x-p'x'\cos\varphi )+xx'\sin^2\varphi
(p'+(x-x'\cos\varphi )^2/p') ,\\
I_x=A_x \ln (y'+C)+A{x-x'\cos\varphi\over C(y'+C)}
+{B_x y'\over Cp'^2}-By'{x-x'\cos\varphi\over p'^2C^3} -
2By'{x-x'\cos\varphi\over p'^4C} ,\\
I_y=-A/C-B/C^3 ,\\
\Longrightarrow f(x,x',y,\varphi )=-{3 x'^2\over 8\pi x}
[I_x+I_y]_{y'_1}^{y'_2}. \\
\end{array}
$$
\begin{figure}[t]
\centering
\epsfbox[80 420 250 760]{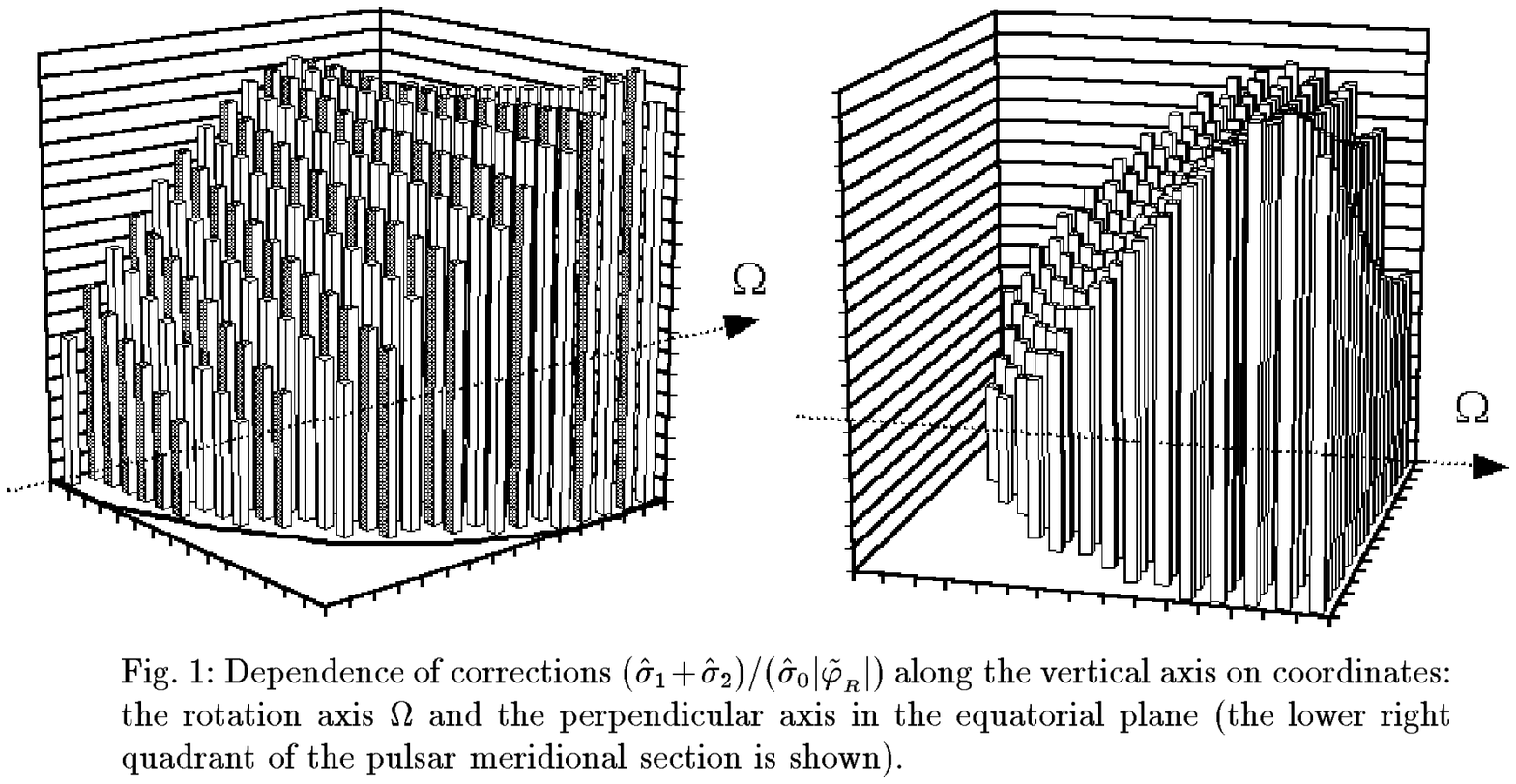}
\label{picture1}
\end{figure}
When deriving the last expression, we took into account 
the fact that the derivative with respect to $y$ could be 
taken inside the integral and that $\partial_y = -\partial_{y'}$.

Integral~(59) can be calculated numerically. The 
integration results with gravielectric corrections~(55) 
are shown in the figure (the maximum amplitude of 
relative corrections is $\approx 1$).

\section{Acknowledgments}
I wish to thank Yu.M. Bruk and A.Yu. Andreev for help. 
I am also grateful to the late D.A. Kirzhnits, with 
whom the ideas and approaches to this study were 
formulated. This work was supported by the Russian 
Foundation for Basic Resreach (project no. 96-15-96616).

Translated by V. Astakhov


\begin{thebibliography}{99}

\bibitem{1}
 A. Yu. Andreev, D. A. Kirzhnits, and S. N. Yudin, 
Pis'ma Zh. Eksp. Teor. Fiz. {\bf 61}, 825 (1995) 
[JETP Lett. {\bf 61}, 846 (1995)].
\bibitem{2}
 I. Bekarevich and I. M. Khalatnikov, 
Zh. Eksp. Teor. Fiz. {\bf 40}, 920 (1961) 
[Sov. Phys. JETP {\bf 13}, 643 (1961)].
\bibitem{3}
 N. N. Bogolyubov and D. V. Shirkov, "Quantum Field" (Nauka, Moscow, 1993).
\bibitem{4}
 A. A. Grib, S. G. Mamaev, and V. M. Mostepanenko, 
"Quantum Effects in Intense External Fields" (Atomizdat, Moscow, 1980).
\bibitem{5}
 J. R. Hartl and D. H. Sharp, Astrophys. J. {\bf 147}, 317 (1967).
\bibitem{6}
 G. B. Hess, Phys. Rev. {\bf 189}, 161 (1967).
\bibitem{7}
 D. A. Kirzhnits and S. N. Yudin, Usp. Fiz. Nauk {\bf 165}, 11 (1995).
\bibitem{8}
 L. D. Landau and E. M. Lifshitz, 
"Fluid Mechanics" (Nauka, Moscow, 1988; Pergamon, Oxford, 1987).
\bibitem{9}
 L. D. Landau and E. M. Lifshitz, 
"The Classical Theory of Fields" (Nauka, Moscow, 1988; Pergamon, Oxford, 1975).
\bibitem{10}
 D. Langlois, astro-ph/0008161.
\bibitem{11}
 R. N. Manchester and J. H. Taylor, "Pulsars" 
(Freeman, San Francisco, 1977; Mir, Moscow, 1980).
\bibitem{12}
 R. Prix, Phys. Rev. D {\bf 62}, 103005 (2000); gr-qc/0004076.

\end{thebibliography}
\end{document}